\documentclass[nofootinbib,showpacs,floatfix,onecolumn,prd,aps,longbibliography]{revtex4-2}
\usepackage{graphicx,psfrag,amsmath,amssymb,amsfonts,latexsym,color,epsf,dcolumn,graphpap,braket,mathrsfs}
\usepackage{caption,subcaption}
\usepackage{float}
\usepackage{lipsum}
\usepackage{enumerate}
\usepackage{cases}
\usepackage{mathtools}
\usepackage{amssymb}

\DeclarePairedDelimiter{\abs}{\lvert}{\rvert}

\usepackage{hyperref}
\hypersetup{colorlinks=true,linktoc=page,linkcolor=blue,citecolor=red,urlcolor=cyan}

\usepackage{tikz-cd} 
\usepackage{tikz}
\usetikzlibrary{shapes.geometric, arrows}

\definecolor{red}{rgb}{1,0,0}
\definecolor{blue}{rgb}{0,0,1}
\definecolor{skyblue}{rgb}{0,0,.5}
\definecolor{green}{rgb}{0,1,0}
\definecolor{orange}{cmyk}{0,.4,1,0}

\usepackage{setspace}

\begin{document}
\onehalfspacing
\title{Nonlocal stress-energy tensor in time-dependent gravitational backgrounds}
\author{Andr\'es Boasso}
\author{Francisco D. Mazzitelli}
\affiliation{Centro At\'omico Bariloche,
Comisi\'on Nacional de Energ\'\i a At\'omica, 
R8402AGP Bariloche, Argentina}
\affiliation{Instituto Balseiro, 
Universidad Nacional de Cuyo,
Centro Atómico Bariloche, 
R8402AGP Bariloche, Argentina}

\begin{abstract}
We analyze the renormalized stress-energy tensor (RSET) of a massless quantum scalar field in time-dependent gravitational backgrounds. Starting from its formal expression obtained within the covariant perturbative expansion to lowest order in the curvature, we evaluate the RSET in an arbitrary number of dimensions in terms of coordinate-space distributions. For time-dependent spherically symmetric spacetimes, we derive a multipole expansion and determine its asymptotic behavior. We find that the RSET is locally nonvanishing at null infinity and depends on the detailed dynamics of the collapsing body. However, the total emitted energy vanishes at this order, meaning that the leading contribution does not account for the energy density of the created particles. Nevertheless, by enforcing stress-tensor conservation up to second order in the curvature, we show that the total radiated energy can be extracted from the first-order RSET. Finally, we compute the induced quantum corrections to the metric at large distances, which display several interesting features.

\end{abstract}

\date{\today}


\maketitle
\section{Introduction}

Quantum field theory in curved spacetime provides the theoretical framework to describe quantum effects of matter fields in the presence of a classical gravitational background. Although gravity itself is not quantized, the presence of quantum fields can lead to important physical phenomena, such as the Hawking radiation of black holes, the generation of cosmological perturbations during inflation, and the appearance of vacuum polarization effects in strong gravitational fields. These processes, first analyzed in the 1970's, revealed that the vacuum of a quantum field depends on the global properties of the spacetime, establishing a deep connection between general relativity, thermodynamics, and quantum theory \cite{birrell1982,Fulling1989,Wald1995,Parker2009,Buchbinder2021}.

In most physically relevant situations, the effects of quantum matter on the background geometry are encoded in the expectation value of the renormalized stress-energy tensor (RSET), which acts as the source term in the semiclassical Einstein equations. The computation of the RSET is, however, a technically challenging problem, since it requires a consistent regularization and renormalization of ultraviolet divergences in a generally covariant way. Exact results can be obtained only for highly symmetric spacetimes, while for generic backgrounds one must rely on approximation schemes.

A particularly powerful and conceptually clear approach is to perform a covariant perturbative expansion of the effective action in powers of the curvature \cite{avramidi1991,Barvinsky1987,Barvinsky1990,Barvinsky1994}.  When one considers as quantum fields the fluctuations around a background metric, the resulting expression can be interpreted as an effective field theory of quantum gravity at low energies \cite{Donoghue1994}. The quantum corrections to the action appear as higher-order curvature terms, which are in general nonlocal and are suppressed by powers of the Planck scale. This perspective, provides a systematic way to compute quantum corrections to classical gravitational phenomena without assuming a full theory of quantum gravity. Within this framework, the nonlocal part of the effective action captures the long-distance propagation of quantum fluctuations, leading to nontrivial effects such as $1/r^3$ corrections to Newton’s potential \cite{Duff1973,Donoghue1993,Dalvit:1994} and running of gravitational couplings \cite{Dalvit1994_2,Barvinsky2023}.

It is therefore of both technical and conceptual interest to compute the RSET within this curvature expansion. Even though this approximation breaks down near strong-field regions such as horizons or singularities, it can still provide valuable insights into the structure of semiclassical gravity in weak and time-dependent backgrounds. In particular, it allows one to study how nonlocality and causality are manifested in the semiclassical stress tensor, and to analyze effects such as vacuum polarization, quantum radiation, and the possible existence of quantum hair in dynamical configurations.

In this work we compute the RSET for a massless scalar field in a generic weakly curved spacetime, starting from the nonlocal effective action at quadratic order in the curvature. We obtain explicit coordinate-space representations for the nonlocal operators involved, and we apply these results to spherically symmetric, time-dependent geometries. This enables us to study in detail the large-distance behavior of the RSET, the possible existence of quantum hair, and the relation between the RSET and particle creation in the semiclassical regime.

The paper is organized as follows. In Sec.~\ref{intro} we briefly review the expression of the RSET derived from the covariant perturbative expansion of the effective action. In Sec.~\ref{sec3} we obtain explicit coordinate-space representations for the nonlocal operators that appear in this expansion. In Sec.~\ref{sec4} we apply these results to time-dependent, spherically symmetric geometries, deriving a multipole expansion of the RSET and analyzing its leading asymptotic behavior at large distances. We analyze the behavior of the RSET at null infinity and compute the backreaction on the metric. In Sec.~\ref{sec:energy} we show how the total energy of particle creation can be extracted from the first-order RSET by using the conservation law at quadratic order in the curvature. In Sec.~\ref{sec:example} we discuss the particular example of a Newtonian oscillating star. Finally, in Sec.~\ref{sec7} we summarize our conclusions and discuss possible extensions of this work.

\subsection{Conventions}
We use Planck units and adopt the mostly plus convention for the metric in the Lorentzian setup $(-,+,+,\dots)$. We also define the Ricci tensor as $R_{\mu\nu}:=R^{\rho}{}_{\mu\rho\nu}$, the Ricci scalar $R=R^{\mu}{}_{\mu}$, and $g=\abs{\mathrm{det}\,g_{\mu\nu}}$. As usual, Greek indices $(\mu,\nu,\dots)$ run over spacetime indices, while Latin indices $(i,j,\dots)$ correspond just to the spatial sector. For Fourier transforms we adopt the convention
\begin{equation}
    \varphi(p)=\int\mathrm{d}^Dx\,\varphi(x)e^{-i\,\eta_{\mu\nu}p^{\mu}x^{\nu}}\,,
\end{equation}
where $\eta_{\mu\nu}$ denotes the Minkowski metric tensor.

\section{RSET for a scalar quantum field in curved spacetime}\label{intro}
We consider a massless scalar quantum field $\phi$ in a $D$-dimensional spacetime endowed with the metric tensor $g_{\mu\nu}$, whose classical action is given by
\begin{equation}
	S = -\frac{1}{2}\int\mathrm{d}^D x\,\sqrt{g}\left(g^{\mu\nu}\partial_{\mu}\phi\partial_{\nu}\phi + \xi R \phi^2\right),
\end{equation}
where $\xi$ is the nonminimal coupling to the curvature. The calculation of the effective action $W$ associated with this quantum field is a complicated task. It can be carried out using a covariant perturbation theory \cite{avramidi1991,Barvinsky1987,Barvinsky1990}, yielding a nonlocal expansion in powers of the curvature. Up to second order, and after performing a Wick rotation, the renormalized nonlocal part of the \textit{in–out} effective action reads \cite{avramidi1991,nosotros_cotton}
\begin{equation}
	W = -\frac{\sqrt{\pi}\,\pi^{\frac{1+(-1)^{D+1}}{2}}(-1)^{\lfloor\frac{D}{2}\rfloor}}{\Gamma\left(\frac{D+3}{2}\right)2^{D+2}(4\pi)^{D/2}}\int\mathrm{d}^D x\,\sqrt{g} \left[ \left(2(D^2-1)\left(\xi-\xi_D\right)^2 - \frac{D}{4(D-1)}\right)R\beta\left(\Box\right)R + R^{\mu\nu}\beta\left(\Box\right)R_{\mu\nu} \right],
	\label{eff_action}
\end{equation}
where $\lfloor \cdot\rfloor$ denotes the floor function, $\xi_D = \frac{D-2}{4(D-1)}$ is conformal coupling to the curvature, $\Box$ is the flat-space d'Alembertian, and $\beta\left(\Box\right)$ a nonlocal operator defined as
\begin{equation}
	\beta\left(\Box\right) =
	\left\{
		\begin{aligned}
			&\left(-\Box\right)^{\frac{D}{2}-2}\qquad&\text{odd }D,\\
			&\left(-\Box\right)^{\frac{D}{2}-2}\log\left(\frac{-\Box}{\mu^2}\right)\qquad&\text{even }D,\\
		\end{aligned}
	\right.
	\label{beta}
\end{equation}
in which the Feynman prescription is implemented.

We are interested in computing the expectation value $\braket{0_{\text{in}}|T_{\mu\nu}|0_{\text{in}}}$ in the \textit{in}-vacuum state $\ket{0_{\text{in}}}$. This can be achieved by evaluating the \textit{in-in} effective action (see, for instance, \cite{Calzetta2008} and references therein). However, a convenient shortcut consists in replacing the Feynman nonlocal operators $\beta(\Box)$ with their retarded counterparts in the expressions for the RSET obtained by varying the \textit{in–out} effective action with respect to the metric \cite{Barvinsky1987,Jordan1987,Campos1993}, 
\begin{equation}
	\braket{T_{\mu\nu}} := \braket{0_{\text{in}}|T_{\mu\nu}|0_{\text{in}}} =  \left.-\frac{2}{\sqrt{g}} \,\frac{\delta W}{\delta g^{\mu\nu}}\right|_{\text{retarded prescription}}\,.
    \label{Tmunu_def}
\end{equation}
We recall that the retarded prescription can be implemented in some operator $Q$ acting on a test function $\varphi(x)$ through its Fourier representation:
\begin{equation}
    Q(-\Box)\varphi(x) = \int\frac{\mathrm{d}^{D}p}{(2\pi)^D}\,Q\left(-(p^0+i\epsilon)^2+|\vec{p}|^2\right)\,\varphi(p)\,e^{i\eta_{\mu\nu}p^{\mu}x^{\nu}},
    \label{retarded}
\end{equation}
with $\epsilon\to 0^+$. We will discuss in detail the structure of  the relevant operators in the next section.

Neglecting the $\mathcal{O}\left(R^2\right)$ terms, Eq.~\eqref{Tmunu_def} yields
\begin{equation}
    \braket{T_{\mu\nu}} = -\frac{\sqrt{\pi}\,\pi^{\frac{1+(-1)^{D+1}}{2}}(-1)^{\lfloor\frac{D}{2}\rfloor}}{\Gamma\left(\frac{D+3}{2}\right)2^{D+2}(4\pi)^{D/2}} \left[ \left(2(D^2-1)\left(\xi-\xi_D\right)^2 - \frac{D}{4(D-1)}\right) \beta\left(\Box\right)H^{(1)}_{\mu\nu} + \beta\left(\Box\right)H^{(2)}_{\mu\nu} \right],
    \label{Tmunu}
\end{equation}
where the tensors $H^{(1)}_{\mu\nu}$ and $H^{(2)}_{\mu\nu}$ are defined as
\begin{equation}
\begin{aligned}
    H^{(1)}_{\mu\nu} &= 4\nabla_{\mu}\nabla_{\nu} R - 4\eta_{\mu\nu}\Box R , \\
    H^{(2)}_{\mu\nu} &= 2\nabla_{\mu}\nabla_{\nu} R - \eta_{\mu\nu}\Box R - 2\Box R_{\mu\nu},
\end{aligned}\label{H1H2}
\end{equation}
with the covariant derivatives (and the d'Alembertian) taken with respect to the flat background, and the curvature scalar evaluated to first order in the weak-field approximation. We emphasize that Eq.~\eqref{Tmunu} is valid provided the retarded prescription in Eq.~\eqref{retarded} is applied to $\beta(\Box)$.

\section{Nonlocal operators as distributions in coordinate space}\label{sec3}

In this section we derive explicit coordinate-space expressions for the nonlocal operators $\beta(\Box)$ that enter the computation of the RSET in Eq.~\eqref{Tmunu}, using the retarded prescription of Eq.~\eqref{retarded}.  
The odd-dimensional case of Eq.~\eqref{beta} can be written as $(-\Box)^{\frac{D-3}{2}}(-\Box)^{-1/2}$, where $\frac{D-3}{2}$ is a non-negative integer.  
Hence, in both even and odd dimensions, $\beta(\Box)$ can be expressed as an non-negative integer power of the d'Alembertian acting on either $\log(-\Box)$ (for even $D$) or $(-\Box)^{-1/2}$ (for odd $D$).  
We therefore focus on analyzing these two basic nonlocal operators.

We begin by considering an arbitrary power of the flat-space d'Alembertian with the retarded prescription in Eq.~\eqref{retarded}, acting on a scalar test function $\varphi$:
\begin{equation}
    (-\Box)^{-\alpha}\varphi(x)
    = \int \frac{\mathrm{d}^D p}{(2\pi)^D}
    \left[-(p^0+i\epsilon)^2 + \abs{\vec{p}}^2\right]^{-\alpha}
    e^{i\eta_{\mu\nu}p^{\mu}x^{\nu}}\varphi(p)\,,
    \label{box_alpha}
\end{equation}
where the limit $\epsilon \to 0^+$ is taken at the end.  
Using the identity from Ref.~\cite{lopez_nacir_fdm},
\begin{equation}
    (p^2)^{-\alpha}
    = \frac{2\sin(\pi\alpha)}{\pi}
      \int_0^\infty \! \mathrm{d}m \, m^{1-2\alpha}
      \frac{1}{p^2+m^2}\,, \qquad 0 < \alpha < 1\,,
\end{equation}
we rewrite Eq.~\eqref{box_alpha} as
\begin{equation}
    (-\Box)^{-\alpha}\varphi(x)
    = \frac{2\sin(\pi\alpha)}{\pi}
      \int \! \mathrm{d}^Dx' \!
      \int_0^\infty \! \mathrm{d}m \, m^{1-2\alpha}
      \underbrace{\int \frac{\mathrm{d}^Dp}{(2\pi)^D}
      \frac{e^{i\eta_{\mu\nu}p^{\mu}(x^{\nu}-{x'}{}^{\nu})}}{-(p^0+i\epsilon)^2+\abs{\vec{p}}^2+m^2}}_{=-G_{\text{R}}(x,x')}
      \varphi(x')\,,
    \label{box_alpha1}
\end{equation}
where $G_{\text{R}}(x,x')$ is the retarded Green function of a free, massive scalar field in flat spacetime. This propagator can be expressed in terms of the Feynman propagator as
\begin{equation}
    G_{\text{R}}(x,x') = 2\Theta(t-t')\,\mathrm{Re}\left[ G_{\text{F}}(x,x')\right]\,,
    \label{Feynman_retarded}
\end{equation}
where $\Theta$ is the Heaviside step function, and~\cite{birrell1982}
\begin{equation}
    G_{\text{F}}(x,x')
    = -\frac{i\pi}{(4\pi i)^{D/2}}
      \left(\frac{4m^2}{\sigma+i0}\right)^{\!\frac{D-2}{4}}
      \! \mathrm{H}^{(2)}_{\frac{D-2}{2}}\!\left(\sqrt{-m^2(\sigma+i0)}\right)\,,
    \label{Feynman_Birrell}
\end{equation}
with $\sigma = \eta_{\mu\nu}(x^\mu - x'^\mu)(x^\nu - x'^\nu) = \abs{\vec{x}-\vec{x}'}^2 - (t-t')^2$ and $\mathrm{H}^{(2)}_\nu$ the Hankel function of the second kind of order $\nu$.

Substituting Eqs.~\eqref{Feynman_retarded}–\eqref{Feynman_Birrell} into Eq.~\eqref{box_alpha1}, and defining $\omega = m\sqrt{-\sigma - i0}$, we obtain
\begin{equation}
    (-\Box)^{-\alpha}\varphi(x)
    = \frac{2\sin(\pi\alpha)}{(2\pi)^{D/2}}
      \int \mathrm{d}^Dx'\, \Theta(t-t')\,
      \mathrm{Re}\!\left[
      \frac{i}{(\sigma+i0)^{D/2-\alpha}}
      \int_0^\infty \! \mathrm{d}\omega\, \omega^{D/2-2\alpha}
      (-i)^{D/2} \mathrm{H}^{(2)}_{\frac{D-2}{2}}(-i\omega)
      \right] \!\varphi(x')\,,
    \label{box_alpha1_0}
\end{equation}
where the inner integral contributes only a numerical factor depending on $D$ and $\alpha$.  
Evaluating it analytically~\cite{gradshteyn2007} yields
\begin{equation}
    (-\Box)^{-\alpha}\varphi(x)
    = -\frac{\sin(\pi\alpha)}{\pi^{D/2+1}}
      2^{1-2\alpha}
      \Gamma\!\left(\tfrac{D}{2}-\alpha\right)
      \Gamma(1-\alpha)
      \int \mathrm{d}^Dx'\,
      \Theta(t-t')\, \mathrm{Im}\!\left[
      (\sigma+i0)^{\alpha - D/2}
      \right]\varphi(x')\,.
    \label{box_alpha1_1}
\end{equation}

To proceed, we use results from distribution theory~\cite{gelfand1964}.  
When $\alpha - D/2$ is not an integer,
\begin{equation}
    (\sigma+i0)^\kappa
    = \Theta(\sigma)\sigma^\kappa
      + e^{i\pi\kappa}\Theta(-\sigma)\abs{\sigma}^\kappa\,,
      \qquad -\kappa \notin \mathbb{N}\,.
    \label{distribution_not_int}
\end{equation}
Inserting this into Eq.~\eqref{box_alpha1_1} gives
\begin{equation}
    (-\Box)^{-\alpha}\varphi(x)
    = -\frac{\sin(\pi\alpha)}{\pi^{D/2+1}}
      2^{1-2\alpha}
      \Gamma\!\left(\tfrac{D}{2}-\alpha\right)
      \Gamma(1-\alpha)
      \sin\!\left[\pi\!\left(\alpha-\tfrac{D}{2}\right)\right]
      \!\int \mathrm{d}^Dx'\,
      \Theta(t-t')\Theta(-\sigma)
      (-\sigma)^{\alpha-D/2}\varphi(x')\,,
    \label{box_alpha1_non_integer}
\end{equation}
which shows that the distribution $(-\Box)^{-\alpha}$ is supported inside the past light cone due to the factor $\Theta(t-t')\Theta(-\sigma)$.  

When $\alpha - D/2 = -n$ with $n \in \mathbb{N}$, one instead has~\cite{gelfand1964}
\begin{equation}
    (\sigma + i0)^{-n}
    = \sigma^{-n}
      + i\,\frac{(-1)^n\pi}{(n-1)!}\,\delta^{(n-1)}(\sigma)\,,
\end{equation}
where $\delta^{(m)}$ is the $m$th derivative of the Dirac delta, and Eq.~\eqref{box_alpha1_1} becomes
\begin{equation}
    (-\Box)^{-\alpha}\varphi(x)
    = 2^{1-2\alpha}\pi^{-D/2}\sin(\pi\alpha)
      (-1)^{D/2 - \alpha - 1}\Gamma(1-\alpha)
      \!\int \mathrm{d}^Dx'\,
      \Theta(t-t')\delta^{(D/2 - \alpha - 1)}(\sigma)\varphi(x')\,.
    \label{box_alpha1_integer}
\end{equation}
The factor $\Theta(t-t')\delta^{(D/2-\alpha-1)}(\sigma)$ restricts the support of the distribution to the surface of the past light cone, thus satisfying Huygens’ principle~\cite{bollini-giambiagi}.

The case $\alpha - D/2 = -n$ is the relevant one for computing the RSET in both even and odd dimensions.  
In odd $D$, $\beta(\Box)$ can be written as an integer power of $\Box$ acting on $(-\Box)^{-1/2}$, hence $\alpha = 1/2$ and $\alpha - D/2 = -n$.  
In even $D$, $\beta(\Box)$ involves and integer power of $\Box$ times $\log(-\Box)$, and this last distribution can be obtained  from Eq.~\eqref{box_alpha1_integer} in the limit $\alpha \to 0$, \footnote{There is a subtle point here. As Eq.~\eqref{box_alpha1_integer} is valid only when $D/2-\alpha-1=-n$, the limit should be taken by varying at the same time $\alpha$ and $D$, while keeping $n$ constant. } recalling that
\begin{equation}
    (-\Box)^{-\alpha} = 1 - \alpha \log(-\Box) + \mathcal{O}(\alpha^2)\,.
\end{equation}
Thus,
\begin{align}
    \log(-\Box)\varphi(x)
    &= \frac{2(-1)^{D/2}}{\pi^{D/2-1}}
       \int \mathrm{d}^Dx'\,
       \Theta(t-t')\,
       \delta^{(D/2-1)}\!\left(\abs{\vec{x}-\vec{x}'}^2 - (t-t')^2\right)\varphi(x')\,,
       && \text{even } D; \label{non-local0}\\[1ex]
    (-\Box)^{-1/2}\varphi(x)
    &= \frac{(-1)^{(D-3)/2}}{\pi^{(D-1)/2}}
       \int \mathrm{d}^Dx'\,
       \Theta(t-t')\,
       \delta^{((D-3)/2)}\!\left(\abs{\vec{x}-\vec{x}'}^2 - (t-t')^2\right)\varphi(x')\,,
       && \text{odd } D.
       \label{non-local}
\end{align}

Applying the remaining powers of $\Box$, we obtain the general expression for $\beta(\Box)$ valid for all $D \ge 3$:
\begin{equation}
    \beta(\Box)\varphi(x)
    = (-1)^{\lfloor(D-3)/2\rfloor}
      \pi^{\frac{1}{2}((-1)^D+1)}
      \frac{2^{D-3}}{\pi^{D/2}}
      \Gamma\!\left(\tfrac{D}{2}-1\right)
      \!\int \mathrm{d}^Dx'\,
      \Theta(t-t')\,
      \delta^{(D-3)}\!\left(\abs{\vec{x}-\vec{x}'}^2 - (t-t')^2\right)
      \varphi(x'),
    \label{beta0}
\end{equation}
which is supported on the past light cone in all dimensions.  
This result agrees with the odd-dimensional case obtained in Ref.~\cite{bollini-giambiagi} and reproduces the four-dimensional expression derived for static backgrounds in Ref.~\cite{fdm_vacuum_polarization}, as well as the time-dependent generalization in Ref.~\cite{calmet_grav}.

Explicit forms can be found by integrating over $t'$ in Eq.~\eqref{beta0}.  
For example,
\begin{equation}
    \beta(\Box)\varphi(t,\vec{x}) =
    \begin{cases}
        \displaystyle
        \frac{1}{2\pi}\int \mathrm{d}^2\vec{x}'\,
        \frac{\varphi}{\abs{\vec{x}-\vec{x}'}}\,, & D=3, \\[1ex]
        \displaystyle
        -\frac{1}{2\pi}\int \mathrm{d}^3\vec{x}'\!
        \left(\frac{\varphi}{\abs{\vec{x}-\vec{x}'}^3}
        + \frac{\dot{\varphi}}{\abs{\vec{x}-\vec{x}'}^2}\right)\,, & D=4, \\[1ex]
        \displaystyle
        -\frac{1}{4\pi^2}\int \mathrm{d}^4\vec{x}'\!
        \left(\frac{3\varphi}{\abs{\vec{x}-\vec{x}'}^5}
        + \frac{3\dot{\varphi}}{\abs{\vec{x}-\vec{x}'}^4}
        + \frac{\ddot{\varphi}}{\abs{\vec{x}-\vec{x}'}^3}\right)\,, & D=5, \\[1ex]
        \displaystyle
        \frac{1}{2\pi^2}\int \mathrm{d}^5\vec{x}'\!
        \left(\frac{15\varphi}{\abs{\vec{x}-\vec{x}'}^7}
        + \frac{15\dot{\varphi}}{\abs{\vec{x}-\vec{x}'}^6}
        + \frac{6\ddot{\varphi}}{\abs{\vec{x}-\vec{x}'}^5}
        + \frac{\dddot{\varphi}}{\abs{\vec{x}-\vec{x}'}^4}\right)\,, & D=6,
    \end{cases}
    \label{examples}
\end{equation}
where dots denote time derivatives and all functions under the integrals are evaluated at $(t-\abs{\vec{x}-\vec{x}'},\vec{x}')$.  
Except for $D=3$, $\beta(\Box)\varphi(x)$ diverges as $\vec{x}\to\vec{x}'$ when $\varphi$ or its derivatives are nonvanishing, but these divergences can be removed in the renormalization procedure.  
Indeed, expanding the numerators around $\vec x'=\vec x$ and introducing a short-distance cutoff $\epsilon>0$, one finds
\begin{equation}
    \beta(\Box)_{\mathrm{div}}\varphi(t,\vec{x}) =
    \begin{cases}
        0\,, & D=3, \\[1ex]
        2\log(\epsilon)\,\varphi(t,\vec{x})\,, & D=4, \\[1ex]
        -\dfrac{3}{2\epsilon}\,\varphi(t,\vec{x})\,, & D=5, \\[1ex]
        \dfrac{10}{\epsilon^2}\,\varphi(t,\vec{x})
        - 2\log(\epsilon)\,\Box\varphi(t,\vec{x})\,, & D=6,
    \end{cases}
    \label{examples_div}
\end{equation}
as $\epsilon\to 0$.  
These divergences are local and covariant, and thus can be absorbed into the bare constants of the theory. 
We emphasize that, in all dimensions, the leading large-distance term in Eq.~\eqref{examples} (lowest power in the denominator) is finite and therefore renormalization-scheme independent.

Finally, although Eq.~\eqref{beta0} was derived for a scalar test function, the result extends straightforwardly to tensor fields in Cartesian coordinates:
\begin{multline}
    \beta(\Box)\varphi_{\mu_1\mu_2\cdots}{}^{\nu_1\nu_2\cdots}(x)
    = (-1)^{\lfloor(D-3)/2\rfloor}
      \pi^{\frac{1}{2}((-1)^D+1)}
      \frac{2^{D-3}}{\pi^{D/2}}
      \Gamma\!\left(\tfrac{D}{2}-1\right) \\
      \times \int \mathrm{d}^Dx'\,
      \Theta(t-t')\,\delta^{(D-3)}\!\left(\abs{\vec{x}-\vec{x}'}^2 - (t-t')^2\right)
      \varphi_{\mu_1\mu_2\cdots}{}^{\nu_1\nu_2\cdots}(x').
    \label{beta0_tensor}
\end{multline}

\section{Multipole expansion of the RSET}\label{sec4}

We now apply Eq.~\eqref{beta0} to spherically symmetric configurations. 
In subsection~\ref{4A}, we evaluate $\beta(\Box)\varphi(x)$ in regions where the test function vanishes, 
obtaining a multipole expansion of the leading-order term of the nonlocal operator. 
As will become clear, this corresponds to expanding the term proportional to $1/r^{D-2}$ 
in temporal derivatives of the test function. 
Subsection~\ref{4B} focuses on the special case of four dimensions, 
for which a closed-form expression can be derived for each term in the asymptotic expansion in powers of $1/r$. 
Finally, in Sec.~\ref{sec:metric} we compute the corresponding quantum corrections to the metric in the large-$r$ limit.

\subsection{Leading order of the RSET in time-dependent scenarios}\label{4A}
We compute the leading term of the RSET, corresponding to the smallest  power of $1/r$. Starting from Eq.~\eqref{beta0},  we handle derivatives of the delta function using
\begin{equation}\begin{aligned}
    \delta^{(n)}(f(t'))g(t') &= (-1)^{n}\delta(f(t'))\underbrace{\frac{\mathrm{d}}{\mathrm{d}t'}\left(\frac{1}{f'(t')}\frac{\mathrm{d}}{\mathrm{d}t'}\left(\frac{1}{f'(t')}\cdots\frac{\mathrm{d}}{\mathrm{d}t'}\left(\frac{g(t')}{f'(t')}\right)\right)\right)}_{n\text{ derivatives}} \\
    &=(-1)^{n}\delta(f(t'))\frac{g^{(n)}(t')}{f'(t')^{n}} + \mathcal{O}\left(\frac{1}{f'(t')^{n+1}}\right),
\end{aligned}\end{equation}
which, for $f(t')=\sigma(t')=\left|\vec{x}-\vec{x}'\right|^2 - \left(t-t'\right)^2$ and $g(t')=\Theta(t-t')\varphi(t',\vec{x}')$ gives
\begin{equation}
    \Theta(t-t') \delta^{\left(D-3\right)}(\sigma) \varphi(t',\vec{x}') = \frac{(-1)^{D-3}}{2^{D-3}} \Theta(t-t') \delta(\sigma)\frac{\partial_{t'}^{D-3}\varphi(t',\vec{x}')}{|\vec{x}-\vec{x}'|^{D-3}} + \mathcal{O}\left( \frac{1}{|\vec{x}-\vec{x}'|^{D-2}} \right).
\end{equation}

After integrating the delta function, Eq.~\eqref{beta0} reduces to
\begin{equation}
    \beta\left(\Box\right) \varphi(x) = \frac{(-1)^{\lfloor\frac{D+2}{2}\rfloor}}{2}\,\pi^{\frac{1}{2}((-1)^D+1)-\frac{D}{2}}\,\Gamma\left(\frac{D}{2}-1\right) \int\mathrm{d}^{D-1}\vec{x}'\left[ \frac{\partial_{t}^{D-3}\varphi(t-|\vec{x}-\vec{x}'|,\vec{x}')}{|\vec{x}-\vec{x}'|^{D-2}} + \mathcal{O}\left( \frac{1}{|\vec{x}-\vec{x}'|^{D-2}} \right)\right],
    \label{beta_multipole1}
\end{equation}
in agreement with the examples provided in Eq.~\eqref{examples}.

Assuming spherical symmetry, $\varphi(t',\vec{x}')=\varphi(t',r')$, and expanding the numerator in a Taylor series yields
\begin{equation}
    \partial_{t}^{D-3}\varphi(t-|\vec{x}-\vec{x}'|,r')=\sum_{n=0}^{\infty}\frac{1}{n!}\partial_{t}^{D-3+n}\varphi(t-r,r')\left(r'\cos\theta'\right)^{n} +\mathcal{O}\left(\frac{1}{r}\right).
\end{equation}
Performing the angular integrals, we obtain
\begin{equation}
    \beta\left(\Box\right) \varphi(x) = (-1)^{\lfloor\frac{D+2}{2}\rfloor}\,\pi^{\frac{(-1)^D}{2}}\,\Gamma\left(\frac{D}{2}-1\right) \frac{1}{r^{D-2}}\sum_{n=0}^{\infty}\frac{1}{4^n n!\Gamma\left(n+\frac{D-1}{2}\right)}\mathcal{M}_n^{(\varphi)}(t-r) +\mathcal{O}\left(\frac{1}{r^{D-1}}\right),
    \label{beta_multipole2}
\end{equation}
where the ``multipole moments'' are defined as
\begin{equation}
    \mathcal{M}_n^{(\varphi)}(t-r)=\int_{0}^{\infty}\mathrm{d}r'\,{r'}^{2n}\,\partial_{t}^{D-3+2n}\varphi(t-r,r').
    \label{multipolar_momenta}
\end{equation}

The leading-order term of the RSET in Eq.~\eqref{Tmunu} follows directly from Eq.~\eqref{beta_multipole2}:

\begin{multline}
    -\braket{T^{t}{}_{t}}=\braket{T^{r}{}_{r}}=\braket{T^{t}{}_{r}} = \frac{1}{r^{D-2}}\, \frac{\pi}{2^{D+2}(4\pi)^{D/2}} \,\frac{\Gamma\left(\frac{D-2}{2}\right)}{\Gamma\left(\frac{D+3}{2}\right)} \\
    \times\left[8(D^{2}-1)(\xi-\xi_{D})^2+\frac{D-2}{D-1}\right]\sum_{n=0}^{\infty}\frac{1}{4^n n!\Gamma\left(n+\frac{D-1}{2}\right)}{\mathcal{{M}}^{(R)}_n}{}''(t-r) +\mathcal{O}\left(\frac{1}{r^{D-1}}\right).
    \label{T_munu_leading}
\end{multline} 
Here ${\mathcal{{M}}^{(R)}_n}{}''$ denotes the second derivative of the multipole moment of order $n$ associated with the curvature scalar $R$, so the leading-order term of the RSET vanishes when the curvature scalar is identically zero throughout spacetime, or when it is time-independent. One can check that Eq.~\eqref{T_munu_leading} satisfies the conservation law $\nabla_{\mu}\braket{T^{\mu}{}_{\nu}}=0$, that in this case has two non-trivial equations:
\begin{equation}\begin{aligned}
    \partial_t \braket{T^{t}{}_{t}} - \partial_r \braket{T^{t}{}_{r}} - \frac{D-2}{r} \braket{T^{t}{}_{r}} &= 0 \\
    \partial_t \braket{T^{t}{}_{r}} + \partial_r \braket{T^{r}{}_{r}} + \frac{D-2}{r} \braket{T^{r}{}_{r}} &= 0.
\end{aligned}\end{equation}

The expansion we carried out in Eq.~\eqref{T_munu_leading} is analogous to the multipole expansion of the radiation in classical electromagnetism, where the different terms of the series correspond to electric dipole, magnetic dipole, electric quadrupole radiation, and so on.

In $D$ dimensions, the leading-order term of the RSET decays asymptotically as $1/r^{D-2}$. Therefore, when computing the flux of radiation through a sphere of radius $r_0\to\infty$, only this term contributes to the flux. Explicitly, the asymptotic flux of energy reads
\begin{multline}
    \int_{r_0\to\infty}\mathrm{d}S_{i}\braket{T^{it}} = \lim_{r_0\to\infty}\frac{2\pi^{\frac{D-2}{2}}}{\Gamma\left(\frac{D-2}{2}\right)}{r_0}^{D-2}\braket{T^{rt}(t-r_0)} \\
    = \frac{1}{2^{2D+1}\Gamma\left(\frac{D+3}{2}\right)} \left[8(D^{2}-1)(\xi-\xi_{D})^2+\frac{D-2}{D-1}\right]\sum_{n=0}^{\infty}\frac{1}{4^n n!\Gamma\left(n+\frac{D-1}{2}\right)}{\mathcal{{M}}^{(R)}_n}''(u),
    \label{energy_flux}
\end{multline}
where $u=t-r_0$ is the coordinate on the future null infinity $\mathscr{I}^{+}$. This non-vanishing energy flux is due, in part, to gravitational pair creation. However, it does not contain the total energy of created particles. Indeed,  as the right-hand side of Eq.~\eqref{energy_flux} is a total derivative, upon integrating over $u\in(-\infty,\infty)$,  the total energy flux in $\mathscr{I}^{+}$ vanishes if the background metric 
is asymptotically flat at both limits $u\to \pm\infty$.

To summarize, we have found that the RSET at linear order in curvature powers carries an asymptotic energy flux, but the total energy in $\mathscr{I}^{+}$ vanishes upon integration over $u$ (this behavior was previously described in Ref.~\cite{quantum_noise}). However, the information of particle  creation is already present in the effective action,  from which the RSET was derived \cite{nosotros_cotton}. In Section \ref{sec:energy} we will clarify this somewhat puzzling situation, showing that  the total energy change in the quantum field due to the time dependent gravitational field can indeed be extracted from the linear-order RSET.

\subsection{The four dimensional case}\label{4B}
In four dimensions, the operator $\log(-\Box/\mu^2)$  acting on a scalar function $\varphi$ reads
\begin{equation}
	\log\left(\frac{-\Box}{\mu^{2}}\right)\varphi(x)=\frac{2}{\pi}\int\mathrm{d}^4x'\, \Theta(t-t') \delta'(\left|\vec{x}-\vec{x}'\right|^2 - \left(t-t'\right)^2)\varphi(x'),
	\label{log_D=4}
\end{equation}
For a spherically symmetry scalar function, and choosing  the integration variables such that $\vec{x}\cdot\vec{x}'=rr'\cos\theta'$, we obtain
\begin{equation}
	\log\left(\frac{-\Box}{\mu^{2}}\right)\varphi(t,r)=4\int_{-\infty}^{t}\mathrm{d}t'\int_{0}^{\infty}\mathrm{d}r'{r'}^{2}\,\varphi(t',r')\int_{-1}^{1}\mathrm{d}(\cos\theta') \,\delta'(r^2+{r'}^2-2rr'\cos\theta'-(t-t')^2).
	\label{log_D=4_1}
\end{equation}

Performing the integrals on $\theta'$ and $t'$ yields
\begin{equation}
    \log\left(\frac{-\Box}{\mu^{2}}\right)\varphi(t,r)=\frac{1}{r}\left( \int_{0}^{\infty}\mathrm{d}r'\frac{r'}{r+r'}\varphi(t-r-r',r') - \int_{0}^{\infty}\mathrm{d}r'\frac{r'}{|r-r'|}\varphi(t-|r-r'|,r') \right).
    \label{log_D=4_3}
\end{equation}
Eq.~\eqref{log_D=4_3} provides an exact formula for the action of $\log(-\Box/\mu^2)$ on any spherically symmetric scalar function, thereby reducing the problem of computing $\log(-\Box/\mu^2)\varphi(t,r)$ to the evaluation of one-dimensional integrals. Moreover, assuming that the test function has compact support,  for large $r$ Eq.~\eqref{log_D=4_3} can be expanded as
\begin{equation}
    \log\left(\frac{-\Box}{\mu^{2}}\right)\varphi(t,r)=\sum_{n=0}^{\infty}\frac{1}{r^{n+2}}M^{(\varphi)}_n(t-r), 
     \label{log_D=4_4}
\end{equation}
where
\begin{equation}
    M^{(\varphi)}_n(t-r)=\int_{0}^{\infty}\mathrm{d}r'\,{r'}^{n+1} \left[ (-1)^{n}\varphi(t-r-r',r')-\varphi(t-r+r',r') \right].
    \label{multipole4D}
\end{equation}
Unlike the  multipolar moments $\mathcal{M}_n^{(\varphi)}$ defined in Eq.~\eqref{multipolar_momenta} for the computation of the leading-order term, the functions $M^{(\varphi)}_n$ defined here do not involve derivatives of the test function $\varphi$. Indeed, Eq.~\eqref{log_D=4_4} provides an all-order asymptotic expansion, involving all powers of $1/r$.

Eq.~\eqref{log_D=4_4} allows us to write down a closed expression for the leading-order term of the RSET in $D=4$,
\begin{equation}
    -\braket{T^{t}{}_{t}} = \braket{T^{t}{}_{r}} = \braket{T^{r}{}_{r}} = \frac{1}{r^2}\frac{1}{16\pi^2}\left[\left(\xi-\frac{1}{6}\right)^2+\frac{1}{180}\right] \int_{0}^{\infty}\mathrm{d}r'\,{r'} \left[ \ddot{R}(t-r+r',r') - \ddot{R}(t-r-r',r') \right] + \mathcal{O}\left(\frac{1}{r^3}\right).
    \label{RSET_4D}
\end{equation}

This result highlights the inherently nonlocal character of the renormalized stress–energy tensor (RSET): a localized source of curvature gives rise to a nontrivial stress–energy distribution of the quantum field throughout spacetime. In contrast with the time-independent case \cite{fdm_vacuum_polarization, calmet_grav, universality_fdm, universality_anderson}, the leading-order term in Eq.~\eqref{RSET_4D} depends explicitly on both the spatial profile and the temporal history of the curvature source, rather than solely on its total associated mass. An explicit example will be analyzed in Sec.~\ref{sec:example}. This dependence of the RSET—and consequently of the quantum-corrected metric—on the detailed spatial structure of the source is often referred to as quantum hair.

\begin{figure}[H]
    \centering
    \includegraphics[width=0.35\linewidth]{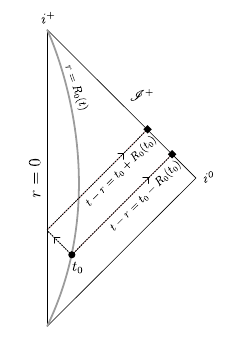}
    \caption{\centering
    Penrose diagram of the quasi-flat spacetime considered in the weak-field approximation. The solid gray line represents a spherical shell with a time-dependent radius $R_0(t)$. A perturbation originating at time $t_0$ (circle mark) modifies the RSET according to Eq.~\eqref{RSET_4D}. The resulting disturbance propagates at the speed of light along two null geodesics $t - r = t_0 \pm R_0(t_0)$ (dashed lines), reaching the future null infinity $\mathscr{I}^+$ (square markers). These two trajectories correspond respectively to an outgoing light ray traveling from the shell toward the exterior region (right), and to an ingoing perturbation emitted from the opposite side of the shell and directed toward the center (left), which continues outward after reaching $r = 0$ and eventually also arrives at $\mathscr{I}^+$.
    }
\label{fig:penrose}
\end{figure}

To gain physical insight into the structure of Eq.~\eqref{RSET_4D}, we consider a simple yet illustrative example: 
a spherical shell with a time-dependent radius $R_0(t)$. 
The corresponding Penrose diagram is shown in Fig.~\ref{fig:penrose}. 
Perturbations produced simultaneously at $t = t_0$ from opposite poles of the shell reach future null infinity $\mathscr{I}^+$ 
at different retarded times, giving rise to the two distinct terms appearing in Eq.~\eqref{RSET_4D}.

\subsection{Quantum corrections to the metric tensor}\label{sec:metric}
In previous studies (see, for instance, Ref.~\cite{fdm_vacuum_polarization}), it was shown that for time-independent backgrounds with compact sources in four dimensions, the RSET falls off as $M/r^5$, with $M$ being the mass characterizing the compact curvature source, leading to quantum corrections to the metric tensor with an asymptotic behavior $M/r^3$. In contrast, in Sec.~\ref{4B} we have shown that, once time dependence is taken into account, the leading-order contribution to the RSET depends on the second time derivative of the curvature scalar and exhibits an asymptotic behavior proportional to ${M''_0}^{(R)}(t-r)/r^2$ (see Eq.~\eqref{RSET_4D}), where $M^{(R)}_0(t-r)$ is the  function defined in Eq.~\eqref{multipole4D} We now examine how this contribution gives rise to  quantum corrections to the metric tensor.

In linearized gravity, writing $g_{\mu\nu} = \eta_{\mu\nu} + h_{\mu\nu}$, the Einstein equations read
\begin{equation}
G_{\mu\nu}^{(1)}=8\pi \left(T_{\mu\nu}^{\textrm{(cl)}}+\langle
   T_{\mu\nu}^{\textrm{(q)}} \rangle\right)\, ,
\label{Einstein}
\end{equation}
where $ G_{\mu\nu}^{(1)}$ denotes the linearized Einstein tensor and $T_{\mu\nu}^{\textrm{(cl)}}$ is the classical part of the stress-energy tensor, which acts as the curvature source of the background. To compute the back-reaction on $h_{\mu\nu}$,  we use the semiclassical gravity approach and  break $h_{\mu\nu}$ into two pieces:
\begin{equation}
    h_{\mu\nu} = h_{\mu\nu}^{\mathrm{(cl)}}+h_{\mu\nu}^{\mathrm{(q)}},
\end{equation}
where $h_{\mu\nu}^{\mathrm{(cl)}}$ denotes the classical part, while $h_{\mu\nu}^{\mathrm{(q)}}$ denotes the quantum one with a similar decomposition for $G_{\mu\nu}^{(1)}$.  Since $h_{\mu\nu}^{\mathrm{(cl)}}$ is the solution of the linearized Einstein equations sourced by the classical stress-energy tensor, then Eq.~\eqref{Einstein} implies that the quantum piece of $G_{\mu\nu}^{(1)}$ satisfies
\begin{equation}
G_{\mu\nu}^{(1,\,\mathrm q)}=8\pi\braket{T_{\mu\nu}}.
    \label{Einstein_q}
\end{equation}

Solving Eq.~\eqref{Einstein_q} requires knowledge of the RSET throughout spacetime. However,  to capture the asymptotic local contribution of $\braket{T_{\mu\nu}}$ to $h_{\mu\nu}^{\mathrm{(q)}}$, and inspired in the Vaidya metric,  we adopt the ansatz
\begin{equation}
-h_{tt}^{\mathrm{(q)}}=h_{tr}^{\mathrm{(q)}}=-h_{rr}^{\mathrm{(q)}} =\frac{1}{r}q(t-r).
\label{ansatz}
\end{equation}
The function $q(t-r)$ in Eq.~\eqref{ansatz} is then determined by substituting the ansatz into Eq.~\eqref{Einstein_q}. 
After a straightforward evaluation of the linearized Einstein tensor for this ansatz one can show that the leading order terms $\sim1/r^2$ match if
\begin{equation}
    q(t-r)=\frac{1}{2\pi}\left[\left(\xi-\frac{1}{6}\right)^2+\frac{1}{180}\right] \int_{0}^{\infty}\mathrm{d}r'\,{r'} \left[ \dot{R}^{\mathrm{(cl)}}(t-r+r',r') - \dot{R}^{\mathrm{(cl)}}(t-r-r',r') \right],
    \label{metric_solution}
\end{equation}
while subleading terms should be fixed by adding subleading contributions to our ansatz. We have added the superscript ${\mathrm{(cl)}}$ to the curvature scalar $R$ in Eq.~\eqref{metric_solution} to emphasize that it must be computed using the classical background metric. 

The complete metric in the asymptotic region $r\to\infty$ outside a spherically-symmetric body of mass $M$ contains a classical part and the quantum correction. We obtain, in spherical coordinates,
\begin{equation}
    h^{\mu}{}_{\nu}=\eta^{\mu\rho}h_{\rho\nu} =  \underbrace{\frac{1}{r}q(t-r)
    \begin{pmatrix}
        1&-1&0&0\\
        1&-1&0&0\\
        0&0&0&0\\
        0&0&0&0
    \end{pmatrix}}_{=h^{\mu}{}_{\nu}^{\mathrm{(q)}}} +
    \underbrace{\frac{2M}{r}
    \begin{pmatrix}
        -1&0&0&0\\
        0&1&0&0\\
        0&0&0&0\\
        0&0&0&0
    \end{pmatrix}}_{=h^{\mu}{}_{\nu}^{\mathrm{(cl)}}} + \mathcal{O}\left(\frac{1}{r^2}\right),
    \label{complete_metric}
\end{equation}
where the function $q$ is given in Eq.~\eqref{metric_solution}. We emphasize that the quantum leading-order term in Eq.~\eqref{complete_metric} is present only for time-dependent backgrounds, and that it is not a vacuum solution of the Einstein equations but is instead sourced by the RSET.

At first glance, Eq.~\eqref{complete_metric} might suggest a quantum-corrected gravitational wave. 
However, several crucial features distinguish it from its classical counterpart. 
First, the propagating mode in Eq.~\eqref{complete_metric} appears even in spherically symmetric configurations, 
whereas classical gravitational waves are absent due to Birkhoff’s theorem. 
Second, the solution does not exhibit the usual transverse polarization. Third, 
the flux of energy produced by the quantum-corrected metric vanishes at large $r$.
Indeed, 
 the gravitational stress-energy pseudo-tensor $t_{\mu\nu}$ is  given by \cite{weinberg},
\begin{equation}
t_{\mu\nu} = \frac{1}{8\pi} \left[
    -\frac{1}{2} h_{\mu\nu} \eta^{\rho\sigma} R^{(1)}_{\rho\sigma}
    + \frac{1}{2} \eta_{\mu\nu} h^{\rho\sigma} R^{(1)}_{\rho\sigma}
    + R^{(2)}_{\mu\nu}
    - \frac{1}{2} \eta_{\mu\nu} \eta^{\rho\sigma} R^{(2)}_{\rho\sigma}
\right] +\mathcal{O}\left(h_{\mu\nu}^3\right)\qquad\text{(Cartesian coordinates)},
\end{equation}
where the superscript $(1)$ denotes the corresponding linearized expression and the superscript $(2)$ the quadratic part in $h_{\mu\nu}=h_{\mu\nu}^{\mathrm{(cl)}}+h_{\mu\nu}^{\mathrm{(q)}}$, a straightforward calculation yields 
$t^{\mu}{}_{\nu}=\mathcal{O}\left(\frac{1}{r^3}\right)\,$. Hence, since $t^{t}{}_{r}$ decays faster than $1/r^2$ as $r\to\infty$, no net energy flux is transported to infinity by the quantum-corrected metric. 
 
As a final step, we derive the geodesic deviation equations for a massive test particle at rest. Given a family of geodesics $x^{\mu}(\tau,s)$ labeled by $s$ and parametrized by the affine parameter $\tau$, we define the deviation vector $S^{\mu}=\mathrm{d}x^{\mu}/\mathrm{d}s$. In the asymptotic region $r\to\infty$, we obtain
\begin{equation}\begin{aligned}
    \frac{\mathrm{d}^2S^{r}}{d\tau^2}&=\frac{2M}{r^3}S^{r}\,,\\
    \frac{\mathrm{d}^2S^{\theta}}{d\tau^2}&=\left(-\frac{M}{r^3}+\frac{\dot{q}(t-r)}{2r^2}\right)S^{\theta}\,,\\
    \frac{\mathrm{d}^2S^{\varphi}}{d\tau^2}&=\left(-\frac{M}{r^3}+\frac{\dot{q}(t-r)}{2r^2}\right)S^{\varphi}\,,
\end{aligned}\end{equation}
where, for consistency, we omitted terms proportional to $q(t-r)/r^3$. The contributions proportional to the source mass $M$ are the well-known results corresponding to the Schwarzschild geometry, while the terms proportional to $\dot q$ in the angular components describe a tidal effect transverse to the propagation direction. Therefore, although the quantum correction to the metric has a polarization tensor that does not represent a genuine gravitational wave, its effect on massive test particles is analog to that of a classical wave.

Finally, we note that, being of order $1/r$, the quantum-corrected metric seems to be  of the same order of magnitude than its classical counterpart at large distances. 
However, restoring conventional units makes the quantum correction proportional to $\ell_{\text{Planck}}^2$, so that  the classical
contribution dominates.
An explicit example  is discussed  in Sec.~\ref{sec:example}.


\section{Energy of created particles from the RSET}\label{sec:energy}
A standard approach to compute the probability of particle creation in semiclassical gravity is to evaluate the imaginary part of the \textit{in–out} effective action $W$, which is related to the vacuum persistence amplitude by
\begin{equation}
    \abs{\braket{0_{\text{out}}|0_{\text{in}}}}^2 = e^{-2\,\mathrm{Im}\, W}\, .
\end{equation}
When the effective action is expanded in powers of the curvature, the leading contribution to $\mathrm{Im}\, W$ is quadratic in the curvature, as shown long ago in Ref.~\cite{Frieman1985} (see also Ref.~\cite{nosotros_cotton} for a more recent discussion).  However, as discussed in the previous section, the integrated flux at future null infinity obtained from the RSET vanishes. At first sight this appears puzzling, since the RSET is derived from the same  effective action.  We  will now show that information about the energy of the created particles can nevertheless be extracted from the RSET computed to first order in the curvature. This can be done by exploiting the conservation law of the full stress–energy tensor, as described in Ref.~\cite{horowitz}.

Let us expand the conservation equation in powers of $h_{\mu\nu}$, treating separately the RSET and the covariant derivative:
\begin{equation}\begin{aligned}
    0 &= \nabla_{\mu}\braket{T^{\mu\nu}} = \left( \nabla_{\mu}^{(0)}+ \nabla_{\mu}^{(1)} + \cdots \right) \left( \braket{T^{\mu\nu}}^{(1)} + \braket{T^{\mu\nu}}^{(2)} +\cdots \right) \\
    &= \underbrace{\nabla_{\mu}^{(0)}\braket{T^{\mu\nu}}^{(1)}}_{=0}+\nabla_{\mu}^{(0)}\braket{T^{\mu\nu}}^{(2)}+\nabla_{\mu}^{(1)}\braket{T^{\mu\nu}}^{(1)} + \mathcal{O}\left((h_{\mu\nu})^3\right)\,,
\end{aligned}\label{conservation}\end{equation}
where the superscript $(n)$ refers to the order in powers of $h_{\mu\nu}$, and $\braket{T^{\mu\nu}}^{(1)}$ is given in Eq.~\eqref{Tmunu}. It is straightforward to verify that the linear-order term in Eq.~\eqref{conservation} vanishes. The quadratic-order term reads
\begin{equation}
\nabla_{\mu}^{(0)}\braket{T^{\mu\nu}}^{(2)}+\underbrace{\nabla_{\mu}^{(1)}\braket{T^{\mu\nu}}^{(1)}}_{:=J^{\nu}} = 0\,,
\label{conservation2}
\end{equation}
and we interpret this equation as a manifestation of energy exchange between the quantum field and the background geometry, encoded through second-order perturbative effects. Let us compute the integral of the vector $J^{\nu}$ over the entire manifold, noting that we must set $\sqrt{g}=1$ to preserve the perturbative order in $h_{\mu\nu}$. In what follows we work in Cartesian coordinates, so that $\nabla_{\mu}^{(0)} = \partial_{\mu}$ and $g_{\mu\nu}^{(0)} = \eta_{\mu\nu}$. The operator $\nabla_{\mu}^{(1)}$ then includes the corresponding linearized Christoffel symbols,
\begin{equation}
    J^{\nu} = \nabla_{\mu}^{(1)}\braket{T^{\mu\nu}}^{(1)} = \Gamma^{\mu}_{\mu\lambda}\braket{T^{\lambda\nu}}^{(1)}+\Gamma^{\nu}_{\mu\lambda}\braket{T^{\mu\lambda}}^{(1)},\qquad
    \Gamma^{\sigma}_{\nu \rho} = \tfrac{1}{2} \eta^{\sigma \lambda} \left( \partial_{\nu} h_{\lambda \rho} + \partial_{\rho} h_{\nu \lambda} - \partial_{\lambda} h_{\nu \rho} \right)\,.
\end{equation}

Using the linearized expressions for the Ricci scalar and tensor, along with the linearized Bianchi identity, $\partial_{\mu}R^{\mu\nu}=\frac{1}{2}\partial^{\nu}R$, after a direct calculation we obtain
\begin{equation}
    \int\mathrm{d}^{D}x\,J^{\nu} = \frac{\sqrt{\pi}\,\pi^{\frac{1+(-1)^{D+1}}{2}}(-1)^{\lfloor\frac{D}{2}\rfloor}}{\Gamma\left(\frac{D+3}{2}\right)2^{D+1}(4\pi)^{D/2}} \int\mathrm{d}^{D}x \left[\partial^{\nu}R^{\mu\lambda}\beta(\Box) R_{\mu\lambda} + \left(2(D^2-1)(\xi-\xi_D)^{2}-\frac{D}{4(D-1)}\right)\partial^{\nu}R\beta(\Box) R \right]\,.
    \label{Jnu}
\end{equation}
To make contact with the particle creation probability, we rewrite Eq.~\eqref{Jnu} in Fourier space:
\begin{multline}
    \int\mathrm{d}^{D}x\,J^{\nu} = \frac{\sqrt{\pi}\,\pi^{\frac{1+(-1)^{D+1}}{2}}(-1)^{\lfloor\frac{D}{2}\rfloor}}{\Gamma\left(\frac{D+3}{2}\right)2^{D+1}(4\pi)^{D/2}} \int\frac{\mathrm{d}^{D}p}{(2\pi)^{D}}\,(-ip^{\nu})\beta_{\text{R}}\left(-p^{2}\right) \\
    \times\left[R^{\mu\lambda}(-p) R_{\mu\lambda}(p) + \left(2(D^2-1)(\xi-\xi_D)^{2}-\frac{D}{4(D-1)}\right)R(-p) R(p) \right]\,,
    \label{Jnu_p}
\end{multline}
where the subscript R indicates that $\beta$ is evaluated with the retarded prescription, which in momentum space reads
\begin{equation}
    \beta_{\text{R}}\left(-p^{2}\right) =
    \left\{
    \begin{aligned}
        &\left(-(p^0+i\epsilon)^2+|\vec{p}|^2\right)^{\frac{D}{2}-2}\,,\qquad&\text{odd }D,\\
        &\left(-(p^0+i\epsilon)^2+|\vec{p}|^2\right)^{\frac{D}{2}-2}\log\left(\frac{-(p^0+i\epsilon)^2+|\vec{p}|^2}{\mu^2}\right)\,,\qquad&\text{even }D,
    \end{aligned}
    \right.
\end{equation}
Taking the limit $\epsilon\to0^{+}$ yields
\begin{equation}
    \beta_{\text{R}}\left(-p^{2}\right) \xrightarrow[\epsilon\to0^{+}]{}
    \left\{
    \begin{aligned}
    &(p^2)^{\frac{D-3}{2}}|p^2|^{-1/2}\left[ \Theta(p^2) + i\Theta(-p^2)\text{sgn}(p^0) \right]\,, \qquad&\text{odd }D\,\\
    &(p^2)^{\frac{D}{2}-2}\left[\log\left|\frac{p^2}{\mu^2}\right|-i\pi\Theta(-p^2)\text{sgn}(p^0)\right]\,,\qquad&\text{even }D.\\
    \end{aligned}
    \right.\label{beta_m}
\end{equation}

Note that the real part of $\beta_{\mathrm{R}}$ in Eq.~\eqref{beta_m} is symmetric under the transformation $p\to-p$, whereas its imaginary part is antisymmetric. The remaining factor in the integrand of Eq.~\eqref{Jnu_p} is also antisymmetric because of $p^{\nu}$. Consequently, only the imaginary part of $\beta_{\mathrm{R}}$ contributes to the integral. Since $R_{\mu\nu}(x)$ is a real function, it follows that $R_{\mu\nu}^{*}(p)= R_{\mu\nu}(-p)$, so the integral is real, as expected. Therefore, Eq.~\eqref{Jnu_p} becomes
\begin{multline}
    \int\mathrm{d}^{D}x\,J^{\nu} = -\frac{\pi^{3/2}(4\pi)^{-D/2}}{\Gamma\left(\frac{D+3}{2}\right)2^{D+1}} \int\frac{\mathrm{d}^{D}p}{(2\pi)^{D}}p^{\nu}\mathrm{sgn}(p^0)\Theta(-p^2)(-p^2)^{\frac{D}{2}-2} \\
    \times\left[R_{\mu\lambda}(p) R^{\mu\lambda}(-p) + \left(2(D^2-1)(\xi-\xi_D)^{2}-\frac{D}{4(D-1)}\right)R(p) R(-p) \right]\,.
    \label{Jnu_p1}
\end{multline}
The integral of $J^{\nu}$ is in general nonzero for all $\nu$. In a spherically symmetric situation the components corresponding to $J^{i}$ vanish due to the symmetry of $R_{\mu\nu}(p)$ under the single-component change $p^{i}\mapsto -p^{i}$. However, the component corresponding to $J^{0}$ remains nonvanishing even in this case.

From Eq.~\eqref{conservation2}, it follows that the component $J^{0}$ encodes precisely the net energy transfer from the background geometry to the quantum excitations of the field:
\begin{equation}
\int\mathrm{d}^{D-1}\vec{x}\left(\left.\braket{T^{00}}^{(2)}\right|_{t\to\infty} - \left.\braket{T^{00}}^{(2)}\right|_{t\to-\infty}\right) + \int_{-\infty}^{\infty}\mathrm{d}t\oint_{\Sigma}\mathrm{d}S_i\braket{T^{i0}}^{(2)} = -\int\mathrm{d}^{D}x\,J^{0}\,.
\label{energy}
\end{equation}
On the left-hand side of Eq.~\eqref{energy}, we can identify two distinct contributions. The first one, which involves the spatial integral of $T^{00}$, corresponds to 
the change in the expectation value of the total energy of the quantum field. The remaining term represents the integrated energy flux through a $(D-2)$-dimensional closed surface $\Sigma$ at $|\vec{x}|\to\infty$, that is, the total radiated energy. Therefore, the left-hand side of Eq.~\eqref{energy} can be interpreted as the total energy change $\Delta E$ of the quantum field produced by the background evolution. Using Eq.~\eqref{Jnu_p1} we obtain
\begin{equation}
    \Delta E = \int\frac{\mathrm{d}^{D}p}{(2\pi)^{D}}\,|p^{0}|\rho(p)\,,
    \label{delta_energy}
\end{equation}
where the function $\rho(p)$, defined by
\begin{equation}
    \rho(p)=\frac{\pi^{\frac{3-D}{2}}}{2^{2D+1}\Gamma\left(\frac{D+3}{2}\right)} \Theta(-p^2)(-p^2)^{\frac{D}{2}-2} \left[R_{\mu\nu}(p) R^{\mu\nu}(-p) + \left(2(D^2-1)(\xi-\xi_D)^{2}-\frac{D}{4(D-1)}\right)R(p) R(-p) \right]\, .
    \label{distribution}
\end{equation}
can be interpreted as the probability density for the creation of a particle with momentum $p$. The factor $\Theta(-p^2)=\Theta((p^0)^2-|\vec{p}|^2)$ in Eq.~\eqref{distribution} represents the usual threshold for the pair creation of massless particles. Only the modes of the background field whose components satisfy $|p^0|>|\vec{p}|$ contribute to the net change in the energy of the quantum field. Notice that, due this constraint, in this approximation energy can be transferred to the quantum field only in the presence of a time-dependent background.

The interpretation of $\rho(p)$ as a probability density becomes more transparent when compared with the explicit computation of the particle creation probability carried out in \cite{nosotros_cotton}, starting from the imaginary part of the effective action at second order in powers of the curvature. It was shown there that the  probability of particle creation, defined as $P=1-\abs{\braket{0_{\mathrm{out}}|0_{\mathrm{in}}}}^2$ is given by
\begin{equation}
    P = \int\frac{\mathrm{d}^{D}p}{(2\pi)^{D}}\,\rho(p)\,.
    \label{proba}
\end{equation}
Importantly, it was also demonstrated in \cite{nosotros_cotton} that the probability density in Eq.~\eqref{distribution} is strictly non-negative, ensuring that the probability of particle creation is well defined and consistent with the physical interpretation of $\Delta E$ as the corresponding energy variation. Alternatively, using the identities presented in \cite{nosotros_cotton}, we can express Eq.~\eqref{distribution} as
\begin{equation}\begin{aligned}
    \rho(p) &= \frac{\pi^{\frac{3-D}{2}}}{4^{D}\Gamma\left(\frac{D+3}{2}\right)} \Theta(-p^2)(-p^2)^{\frac{D}{2}-2} \left[ \left(D^2-1\right)\left(\xi - \xi_D\right)^2  R\left(-p\right)R\left(p\right) + \frac{D-2}{8\left(D-3\right)} C^{\mu\nu\rho\sigma}\left(-p\right)C_{\mu\nu\rho\sigma}\left(p\right) \right], \quad &D\geq 4\\
    &=\frac{\pi^{\frac{3-D}{2}}}{4^{D}\Gamma\left(\frac{D+3}{2}\right)} \Theta(-p^2)(-p^2)^{\frac{D}{2}-2} \left[\left(D^2-1\right)\left(\xi - \xi_D\right)^2 R\left(-p\right)R\left(p\right) + \frac{1}{4p^2} C^{\mu\nu\rho}\left(-p\right)C_{\mu\nu\rho}\left(p\right) \right], &D\geq 3,
    \label{distribution_C}
\end{aligned}\end{equation}
where $C_{\mu\nu\rho\sigma}$ is the linearized Weyl tensor, and $C_{\mu\nu\rho}$ is the linearized expression of the Cotton tensor, defined by
\begin{equation}
    C_{\mu\nu\rho} = \nabla_{\rho}R_{\mu\nu} - \nabla_{\nu}R_{\mu\rho} + \frac{1}{2(D-1)}\left( g_{\mu\rho}\nabla_{\nu}R - g_{\mu\nu}\nabla_{\rho}R \right) .
\end{equation}
The expressions for $\rho(p)$ in terms of the Weyl or Cotton tensor in Eq.~\eqref{distribution_C} show that conformally coupled theories ($\xi=\xi_D$) in conformally flat backgrounds ($C_{\mu\nu\rho}=0$ in $D\geq3$ or $C_{\mu\nu\rho\sigma}=0$ in $D\geq4$) exhibit no particle creation, and therefore no net transfer of energy to the quantum field.

To summarize, in this section we have shown that by enforcing the covariant conservation law 
of the second-order RSET, the total energy of the created particles can be obtained by integrating the first-order RSET contracted with local geometric quantities over the entire manifold. 
The final expression, Eq.~\eqref{delta_energy}, admits a clear physical interpretation 
and could, in fact, have been anticipated from previous analyses 
\cite{Frieman1985, nosotros_cotton}. 
Nevertheless, the derivation presented here makes explicit that this information 
is already encoded in the first-order RSET itself.

\section{Newtonian oscillating star in four dimensions}\label{sec:example}

In this section, we apply our results to an example in four dimensions. We employ the Newtonian oscillating star model proposed in \cite{nosotros_cotton} to compute the corresponding RSET, the backreaction on the metric, and the energy exchange between the background and the quantum field. The line element reads:
\begin{equation}
{\rm d}s^2 = - \big(1+2\Phi(t,r)\big){\rm d}t^2 +\big(1-2\Phi(t,r)\big)({\rm d}\vec x)^2 \, ,
\label{metrica_estrella}
\end{equation}
where $r=|\vec{x}|$, and $\Phi(t,r)$ is the classical Newtonian potential. Outside a spherically symmetric star, $\Phi(r) = -M/r$, where $M$ is the total mass. For a time-dependent radius $a(t)$, it is assumed that $\Phi(t,r)$ depends on $t$ solely locally through a function $a(t)$, and that it interpolates between a constant value at $r=0$ and $-M/a(t)$ at $r=a(t)$, that is 
\begin{align}
    \Phi\left(t,r\right)= \begin{cases}
	-\frac{M}{r}f(r/a(t)) & \text{if } r < a(t) \\
	-\frac{M}{r} & \text{if } r \geq a(t)\,, 
    \end{cases}
\label{ejemplo2:potencialf}
\end{align}
and the function $f$ depends on the structure of the star.

The Newtonian potential is assumed to be continuous at $r=a(t)$, along with its first and second $r$-derivatives to avoid singular terms in the curvature. These continuity conditions imply $f(1)=1$ and $f'(1)=f''(1)=0$, where the primes denote derivatives with respect to the argument. For instance, a possible choice of the function $f$ satisfying the continuity constraints is an odd polynomial of fifth order: $f(x)=\frac{15}{8}x(1-\frac{2}{3}x^2+\frac{1}{5}x^4)$, which has been used in Ref.~\cite{estrella_gas}.

We model small radial oscillations of frequency $\omega$ occurring in a characteristic timescale $\tau$ as
\begin{equation}
    a(t) = a_0 \left[1 + \epsilon\, e^{-\frac{\pi}{2}\frac{t^2}{\tau^2}}\text{cos}(\omega t)\right]\,,
\end{equation}
with $a_0$ constant, and $\epsilon \ll 1$. The Newtonian potential takes the form
\begin{equation}
    \Phi(t,r) = \Phi_0(r) + \epsilon \frac{M}{a_0}f'(r/a_0)\, e^{-\frac{\pi}{2}\frac{t^2}{\tau^2}} \text{cos}(\omega t) \,\Theta\left(a_0-r\right) + \mathcal{O}\left(\epsilon^2\right)\,,
    \label{ejemplo2:pot_epsilon}
\end{equation}
where the time-independent function $\Phi_0$ denotes the potential $\Phi$ of Eq.~\eqref{ejemplo2:potencialf} with $a(t) = a_0$. In what follows, we shall retain terms up to linear order in $\epsilon$. The Ricci tensor corresponding to the metric in Eq.~\eqref{metrica_estrella} up to linear order in $\Phi$ is
\begin{equation}\begin{aligned}
    R_{00} &= \Box\Phi + 4\partial_{0}^{2}\Phi\,, \\
    R_{0i} &= 2\partial_{0}\partial_{i}\Phi\,, \\
    R_{ij} &= \delta_{ij}\Box\Phi\,,
\end{aligned}
    \label{example_scalar}
\end{equation}
where $\Box$ denotes the flat-space d'Alembertian.

The leading-order contribution to the RSET can be readily computed using Eq.~\eqref{RSET_4D}. In the limit where the duration timescale of the oscillations is much greater than their period, and this latter is much greater than the radius of the star ($\tau\gg1/\omega\gg a_0$), we obtain
\begin{equation}
    -\braket{T^{t}{}_{t}} = \braket{T^{t}{}_{r}} = \braket{T^{r}{}_{r}} = \frac{1}{r^2}\frac{1}{2\pi^2}\left[\left(\xi-\frac{1}{6}\right)^2+\frac{1}{180}\right] \epsilon M\omega^5{a_0}^2\alpha_f\, e^{-\frac{\pi}{2}\frac{(t-r)^2}{\tau^2}} \sin\left(\omega(t-r)\right)\,,
    \label{result_example}
\end{equation}
where the structure-dependent constant $\alpha_f$ is defined as in Ref.~\cite{nosotros_cotton} as
\begin{equation}
    \alpha_f = \int_{0}^{1}\mathrm{d}x\,x^2 f'(x)\,.
    \label{alpha_f}
\end{equation}

The leading-order term of the RSET in Eq.~\eqref{result_example} is a Gaussian wave packet with a characteristic size of $\tau$ and a phase frequency of $\omega$. The amplitude is proportional to the total mass of the star and to the constant $\alpha_f$, which encodes the internal structure of the star. In contrast with time-independent scenarios \cite{fdm_vacuum_polarization}, the leading-order term in the RSET encodes information about the internal structure of the star. The asymptotic behavior of the component $\braket{T^{t}{}_{r}}\sim1/r^2$ ensures a non-vanishing energy flux through a closed surface at $r\to\infty$, and it can be readily verified that the total radiated energy vanishes when integrated over time, as expected for the RSET at first order in curvature.

On the other hand, and following the procedure described in Section \ref{sec:metric}, we can compute the quantum corrections to the metric. By substituting the classical curvature scalar from Eq.~\eqref{example_scalar} into Eqs.~\eqref{metric_solution} and \eqref{ansatz}, we obtain the leading-order correction in the form
\begin{equation}
    -h^{t}{}_{t}^{\mathrm{(q)}}=h^{t}{}_{r}^{\mathrm{(q)}}=h^{r}{}_{r}^{\mathrm{(q)}} =
    \frac{4}{\pi\, r}\left[\left(\xi-\frac{1}{6}\right)^2+\frac{1}{180}\right]\epsilon M\omega^4{a_0}^2\alpha_f\, e^{-\frac{\pi}{2}\frac{(t-r)^2}{\tau^2}} \cos\left(\omega(t-r)\right)\,.
\end{equation}
As before, the corrections to the metric depends on the details of the star's interior, and the phenomenon of quantum hair already appears at leading order.

It is interesting to compare this result with the classical metric, which is of order $M/r$. Restoring conventional units, the quantum corrections overwhelm the classical metric when
\begin{equation}
\ell_{\text{Planck}}\gtrsim
\frac{\lambda^2}{a_0}\,,
\end{equation}
where $\lambda$ is the wavelength of the oscillations. This gives unrealistically small values for the ratio $\lambda^2/a_0$. However, the  quantum corrections for the time-dependent situation can be larger than the correction in the time independent case, which in natural units is of order $M/r^3$. Indeed this happens when 
\begin{equation}
r \gtrsim \frac{\lambda^2}{a_0}\,,
\end{equation}
which of course does not depend on the Planck length. 

Finally, we compute the total energy $\Delta E$ transferred to the quantum field during the radial oscillations of the Newtonian star. In order to apply the expression Eq.~\eqref{delta_energy}, we first compute the Fourier transforms of the Ricci scalar and tensor in Eq.~\eqref{example_scalar}, or equivalently, the Fourier transform of the Newtonian potential of Eq.~\eqref{ejemplo2:pot_epsilon}. We obtain
\begin{equation}
    \Phi(p)\approx 2\pi\delta(p^0)\Phi_0(|\vec{p}|) + 2\sqrt{2}\pi\epsilon M \alpha_f \, a_0^2 \tau  \left(e^{-\frac{\tau ^2}{2 \pi }(p^0-\omega )^2}+e^{-\frac{\tau ^2}{2 \pi }(p^0+\omega )^2}\right)\,,
\end{equation}
where the structure-dependent constant $\alpha_f$ is defined in Eq.~\eqref{alpha_f}, and the approximation is valid when $\tau\gg1/\omega\gg a_0$ is considered. A direct calculation of the probability density $\rho(p)$ defined in Eq.~\eqref{distribution} yields
\begin{equation}
    \rho(p)=\pi(\epsilon M \alpha_f)^2 \, a_0^4 \tau^2  \Theta(-p^2) \left[\left(\xi-\frac{1}{6}\right)^{2}\left(|\vec{p}|^2-3 (p^0)^2\right)^2 + \frac{1}{45}|\vec{p}|^4 \right] \left(e^{-\frac{\tau ^2}{2 \pi }(p^0-\omega )^2}+e^{-\frac{\tau ^2}{2 \pi }(p^0+\omega )^2}\right)^2\,,
    \label{lambda_example}
\end{equation}
where the terms proportional to $\delta(p^0)$ vanish due to the presence of the factor $\Theta(-p^2)$. Inserting Eq.~\eqref{lambda_example} in Eq.~\eqref{delta_energy} yields
\begin{equation}
    \Delta E = \frac{34}{35\pi}\left[\left(\xi-\frac{1}{6}\right)^2+\frac{1}{612}\right](\epsilon M\alpha_f)^2{a_0}^4\omega^8\tau ,
\end{equation}
where we have discarded the terms with negative powers of $\omega\tau$ because of the limit $\tau\gg1/\omega$.

\section{Conclusions}\label{sec7}

We summarize here the main new results obtained in this work. We have analyzed in detail the properties of the RSET for massless fields in curved spacetime, focusing on time-dependent situations. Our starting point was the RSET computed up to linear order in the curvature, using a covariant perturbative expansion. This RSET can be expressed in terms of nonlocal operators involving $\log(-\Box)$ in even dimensions and $(-\Box)^{-1/2}$ in odd dimensions. Starting from an integral representation of $(-\Box)^{-\alpha}$ in terms of the massive retarded propagator, we obtained explicit expressions for these nonlocal operators as distributions in configuration space. Despite their different appearance, both $\log(-\Box)$ in even dimensions and $(-\Box)^{-1/2}$ in odd dimensions can be written in terms of derivatives of the Dirac delta function supported on the past light cone. This shows that the corresponding vacuum fluctuations satisfy the Huygens' principle.

We then applied this representation to compute the RSET in time-dependent geometries with spherical symmetry, concentrating on the asymptotic behavior at large distances through a multipole expansion. Our main goal was to investigate the existence of quantum hair, namely, a dependence of the RSET on the internal structure of a collapsing or oscillating star in the weak-field approximation. In previous analyses restricted to static geometries, quantum hair appears only in subleading multipoles
\cite{fdm_vacuum_polarization,Calmet2019,Calmet2021}. In contrast, in time-dependent situations it already arises at leading order, meaning that the quantum hair manifests itself in the flux at null infinity. We obtained explicit expressions in four dimensions, showing that the RSET at null infinity can be written in terms of integrals of derivatives of the Ricci tensor over the entire past history of the source.

Although the flux at infinity is instantaneously nonvanishing, the total emitted energy vanishes. In other words, the RSET computed at first order in the curvature, and evaluated at $\mathscr{I}^+$, does not contain the total energy of the created particles. This fact is well known. However, the effective action from which the RSET is derived contains the information of the probability of pair creation in its imaginary part. Therefore, it should still be possible to extract the total energy of the created particles from the lowest-order RSET. We have shown explicitly that this is indeed the case, following the idea of Ref.~\cite{horowitz}. The key point is that, although the second-order RSET is not known, it must be covariantly conserved. Expanding the conservation law in powers of the curvature reveals that the total energy is encoded in the first-order RSET. The resulting expression is closely related to the vacuum persistence probability, determined by the imaginary part of the effective action.

We also used the four-dimensional RSET to compute the backreaction on the metric. The quantum-corrected metric exhibits some interesting properties: unlike a classical gravitational wave, it does not vanish even under spherical symmetry. Moreover, although the metric components decay as $1/r$, we have verified that the associated stress-energy pseudo-tensor decays faster than \(1/r^2\) as \(r \to \infty\).

Finally, we illustrated some of our results by providing explicit calculations for the case of an oscillating Newtonian star in four dimensions.

Regarding future work, the methods and results presented here could be extended to charged and/or slowly rotating configurations, still within the weak-field approximation. Beyond this regime, a possible avenue is to explore generalizations of the nonlocal effective action applicable to strong gravitational fields. Since this is a technically demanding task (the third-order corrections already display considerable complexity \cite{Barvinsky:1994}) several approaches have been developed to partially incorporate the quantum effects.
For example, long time ago, Vilkovisky and collaborators proposed an effective action for conformal fields in which the nonlocal form factors $\beta(\Box)$
are evaluated on a conformally flat metric rather than on flat spacetime \cite{Fradkin:1978, Vilkovisky:1984}. A local approximation to this effective action has been employed in analyses of gravitational collapse \cite{Frolov:1981}. In a similar direction, various authors have considered nonlinear completions of the second-order effective action, where the nonlocal form factors are evaluated in cosmological backgrounds \cite{Hamber:2005, lopez_nacir_fdm, Donoghue:2014}. Other approaches to go beyond the weak-field regime include performing a spherical dimensional reduction and subsequently applying the  $(1+1)$-dimensional Polyakov effective action \cite{Fabbri:2005}, as well as exploring its generalization to $3+1$ dimensions through the Riegert action \cite{Riegert:1984,Mottola:2006,Shapiro:2008sf}. A systematic comparison and possible extensions of these frameworks are currently underway.

\section* {Acknowledgments} 
This research was supported by  Consejo Nacional de Investigaciones Científicas y Técnicas (CONICET).


\bibliographystyle{apsrev4-2} 
\bibliography{bibliography} 

@article{avramidi1991,
title = {A covariant technique for the calculation of the one-loop effective action},
journal = {Nuclear Physics B},
volume = {355},
number = {3},
pages = {712-754},
year = {1991},
issn = {0550-3213},
doi = {https://doi.org/10.1016/0550-3213(91)90492-G},
url = {https://www.sciencedirect.com/science/article/pii/055032139190492G},
author = {I.G. Avramidi}
}

@book{birrell1982,
    author = "Birrell, N. D. and Davies, P. C. W.",
    title = "{Quantum Fields in Curved Space}",
    doi = "10.1017/CBO9780511622632",
    isbn = "978-0-511-62263-2, 978-0-521-27858-4",
    publisher = "Cambridge University Press",
    address = "Cambridge, UK",
    series = "Cambridge Monographs on Mathematical Physics",
    year = "1982"
}

@book{gelfand1964,
  author    = {I. M. Gelfand and G. E. Shilov},
  title     = {Generalized Functions, Volume 1: Properties and Operations},
  publisher = {AMS Chelsea Publishing},
  year      = {1964},
  volume    = {377},
  isbn      = {978-1-4704-2658-3},
  url       = {https://bookstore.ams.org/chel-377-h}
}

@book{gradshteyn2007,
  author    = {I. S. Gradshteyn and I. M. Ryzhik},
  editor    = {Alan Jeffrey and Daniel Zwillinger},
  title     = {Table of Integrals, Series, and Products},
  publisher = {Elsevier/Academic Press},
  year      = {2007},
  edition   = {7th},
  address   = {Amsterdam},
  isbn      = {978-0-12-373637-6},
  url       = {https://archive.org/details/GradshteinI.S.RyzhikI.M.TablesOfIntegralsSeriesAndProducts}
}

@article{bollini-giambiagi,
	title = {Arbitrary powers of {D}’{Alembertians} and the {Huygens}’ principle},
	volume = {34},
	issn = {0022-2488, 1089-7658},
	url = {https://pubs.aip.org/jmp/article/34/2/610/442024/Arbitrary-powers-of-D-Alembertians-and-the-Huygens},
	doi = {10.1063/1.530263},
	language = {en},
	number = {2},
	urldate = {2023-12-05},
	journal = {Journal of Mathematical Physics},
	author = {Bollini, C. G. and Giambiagi, J. J.},
	month = feb,
	year = {1993},
	pages = {610--621},
}

@article{lopez_nacir_fdm,
	title = {Running of {Newton}’s constant and noninteger powers of the d’{Alembertian}},
	volume = {75},
	issn = {1550-7998, 1550-2368},
	url = {https://link.aps.org/doi/10.1103/PhysRevD.75.024003},
	doi = {10.1103/PhysRevD.75.024003},
	language = {en},
	number = {2},
	urldate = {2023-12-05},
	journal = {Physical Review D},
	author = {López Nacir, D. and Mazzitelli, F. D.},
	month = jan,
	year = {2007},
	pages = {024003},
}

@article{fdm_vacuum_polarization,
	title = {Vacuum polarization around stars: {Nonlocal} approximation},
	volume = {71},
	issn = {1550-7998, 1550-2368},
	shorttitle = {Vacuum polarization around stars},
	url = {https://link.aps.org/doi/10.1103/PhysRevD.71.064001},
	doi = {10.1103/PhysRevD.71.064001},
	language = {en},
	number = {6},
	urldate = {2023-12-05},
	journal = {Physical Review D},
	author = {Satz, Alejandro and Mazzitelli, Francisco D. and Alvarez, Ezequiel},
	month = mar,
	year = {2005},
	pages = {064001},
}

@article{calmet_grav,
	title = {Gravitational radiation in quantum gravity},
	volume = {78},
	issn = {1434-6044, 1434-6052},
	url = {http://link.springer.com/10.1140/epjc/s10052-018-6265-3},
	doi = {10.1140/epjc/s10052-018-6265-3},
	language = {en},
	number = {9},
	urldate = {2023-12-05},
	journal = {The European Physical Journal C},
	author = {Calmet, Xavier and El-Menoufi, Basem Kamal and Latosh, Boris and Mohapatra, Sonali},
	month = sep,
	year = {2018},
	pages = {780},
}

@article{nosotros_cotton,
  title = {Nonlocal effective action and particle creation in $D$ dimensions},
  author = {Boasso, Andr\'es and Franchino Vi\~nas, Sebasti\'an and Mazzitelli, Francisco D.},
  journal = {Phys. Rev. D},
  volume = {111},
  issue = {8},
  pages = {085023},
  numpages = {11},
  year = {2025},
  month = {Apr},
  publisher = {American Physical Society},
  doi = {10.1103/PhysRevD.111.085023},
  url = {https://link.aps.org/doi/10.1103/PhysRevD.111.085023}
}

@article{estrella_gas,
author = {Jun, Jung-Hwan and Kwak, Ho Young},
year = {2012},
month = {01},
pages = {},
title = {Gravitational collapse of Newtonian stars},
volume = {09},
journal = {International Journal of Modern Physics D},
doi = {10.1142/S0218271800000049}
}

@article{horowitz,
  title = {Quantum stress energy in nearly conformally flat spacetimes},
  author = {Horowitz, Gary T. and Wald, Robert M.},
  journal = {Phys. Rev. D},
  volume = {21},
  issue = {6},
  pages = {1462--1465},
  numpages = {0},
  year = {1980},
  month = {Mar},
  publisher = {American Physical Society},
  doi = {10.1103/PhysRevD.21.1462},
  url = {https://link.aps.org/doi/10.1103/PhysRevD.21.1462}
}

@book{weinberg,
    author = "Weinberg, Steven",
    title = "{Gravitation and Cosmology}: {Principles and Applications of the General Theory of Relativity}",
    isbn = "978-0-471-92567-5, 978-0-471-92567-5",
    publisher = "John Wiley and Sons",
    address = "New York",
    year = "1972"
}

@article{universality_fdm,
  title = {Conformal invariance and apparent universality of semiclassical gravity},
  author = {Garbarz, A. and Giribet, G. and Mazzitelli, F. D.},
  journal = {Phys. Rev. D},
  volume = {78},
  issue = {8},
  pages = {084014},
  numpages = {8},
  year = {2008},
  month = {Oct},
  publisher = {American Physical Society},
  doi = {10.1103/PhysRevD.78.084014},
  url = {https://link.aps.org/doi/10.1103/PhysRevD.78.084014}
}

@article{universality_anderson,
  title = {Apparent universality of semiclassical gravity in the far field limit},
  author = {Anderson, Paul R. and Fabbri, A.},
  journal = {Phys. Rev. D},
  volume = {75},
  issue = {4},
  pages = {044015},
  numpages = {6},
  year = {2007},
  month = {Feb},
  publisher = {American Physical Society},
  doi = {10.1103/PhysRevD.75.044015},
  url = {https://link.aps.org/doi/10.1103/PhysRevD.75.044015}
}

@article{quantum_noise,
    author = "Mirzabekian, A. G. and Vilkovisky, G. A.",
    title = "{Vacuum radiation in covariant perturbation theory}",
    doi = "10.1016/S0370-2693(97)01068-X",
    journal = "Phys. Lett. B",
    volume = "414",
    pages = "123--129",
    year = "1997"
}

@book{Fulling1989,
    author = "Fulling, S. A.",
    title = "{Aspects of Quantum Field Theory in Curved Space-time}",
    volume = "17",
    year = "1989"
}

@book{Parker2009,
    author = "Parker, Leonard E. and Toms, D.",
    title = "{Quantum Field Theory in Curved Spacetime}: {Quantized Field and Gravity}",
    doi = "10.1017/CBO9780511813924",
    isbn = "978-0-521-87787-9, 978-0-521-87787-9, 978-0-511-60155-2",
    publisher = "Cambridge University Press",
    series = "Cambridge Monographs on Mathematical Physics",
    month = "8",
    year = "2009"
}

@book{Wald1995,
    author = "Wald, Robert M.",
    title = "{Quantum Field Theory in Curved Space-Time and Black Hole Thermodynamics}",
    isbn = "978-0-226-87027-4",
    publisher = "University of Chicago Press",
    address = "Chicago, IL",
    series = "Chicago Lectures in Physics",
    year = "1995"
}

@book{Buchbinder2021,
    author = "Buchbinder, Iosif L. and Shapiro, Ilya",
    title = "{Introduction to Quantum Field Theory with Applications to Quantum Gravity}",
    doi = "10.1093/oso/9780198838319.001.0001",
    isbn = "978-0-19-887234-4, 978-0-19-883831-9",
    publisher = "Oxford University Press",
    series = "Oxford Graduate Texts",
    month = "3",
    year = "2021"
}

@article{Barvinsky1987,
    author = "Barvinsky, A. O. and Vilkovisky, G. A.",
    title = "{Beyond the Schwinger-Dewitt Technique: Converting Loops Into Trees and In-In Currents}",
    doi = "10.1016/0550-3213(87)90681-X",
    journal = "Nucl. Phys. B",
    volume = "282",
    pages = "163--188",
    year = "1987"
}

@article{Barvinsky2023,
    author = "Barvinsky, A. O. and Wachowski, W.",
    title = "{Notes on conformal anomaly, nonlocal effective action, and the metamorphosis of the running scale}",
    eprint = "2306.03780",
    archivePrefix = "arXiv",
    primaryClass = "hep-th",
    doi = "10.1103/PhysRevD.108.045014",
    journal = "Phys. Rev. D",
    volume = "108",
    number = "4",
    pages = "045014",
    year = "2023"
}

@article{Barvinsky1990,
    author = "Barvinsky, A. O. and Vilkovisky, G. A.",
    title = "{Covariant perturbation theory. 2: Second order in the curvature. General algorithms}",
    doi = "10.1016/0550-3213(90)90047-H",
    journal = "Nucl. Phys. B",
    volume = "333",
    pages = "471--511",
    year = "1990"
}

@article{Barvinsky1994,
    author = "Barvinsky, A. O. and Gusev, Yu. V. and Vilkovisky, G. A. and Zhytnikov, V. V.",
    title = "{The One loop effective action and trace anomaly in four-dimensions}",
    eprint = "hep-th/9404187",
    archivePrefix = "arXiv",
    reportNumber = "ALBERTA-THY-15-94",
    doi = "10.1016/0550-3213(94)00585-3",
    journal = "Nucl. Phys. B",
    volume = "439",
    pages = "561--582",
    year = "1995"
}

@article{Donoghue1994,
    author = "Donoghue, John F.",
    title = "{General relativity as an effective field theory: The leading quantum corrections}",
    eprint = "gr-qc/9405057",
    archivePrefix = "arXiv",
    reportNumber = "UMHEP-408",
    doi = "10.1103/PhysRevD.50.3874",
    journal = "Phys. Rev. D",
    volume = "50",
    pages = "3874--3888",
    year = "1994"
}

@article{Donoghue1993,
    author = "Donoghue, John F.",
    title = "{Leading quantum correction to the Newtonian potential}",
    eprint = "gr-qc/9310024",
    archivePrefix = "arXiv",
    reportNumber = "UMHEP-396",
    doi = "10.1103/PhysRevLett.72.2996",
    journal = "Phys. Rev. Lett.",
    volume = "72",
    pages = "2996--2999",
    year = "1994"
}

@article{Dalvit:1994,
    author = "Dalvit, Diego A. R. and Mazzitelli, Francisco D.",
    title = "{Running coupling constants, Newtonian potential and nonlocalities in the effective action}",
    eprint = "gr-qc/9402003",
    archivePrefix = "arXiv",
    reportNumber = "GTCRG-94-01",
    doi = "10.1103/PhysRevD.50.1001",
    journal = "Phys. Rev. D",
    volume = "50",
    pages = "1001--1009",
    year = "1994"
}

@article{Dalvit1994_2,
    author = "Dalvit, Diego A. R. and Mazzitelli, Francisco D.",
    title = "{Heat kernel and scaling of gravitational constants}",
    eprint = "hep-th/9410032",
    archivePrefix = "arXiv",
    doi = "10.1103/PhysRevD.52.2577",
    journal = "Phys. Rev. D",
    volume = "52",
    pages = "2577--2580",
    year = "1995"
}

@article{Duff1973,
    author = "Duff, M. J.",
    title = "{Quantum Tree Graphs and the Schwarzschild Solution}",
    doi = "10.1103/PhysRevD.7.2317",
    journal = "Phys. Rev. D",
    volume = "7",
    pages = "2317--2326",
    year = "1973"
}

@article{Jordan1987,
    author = "Jordan, R. D.",
    title = "{Stability of Flat Space-time in Quantum Gravity}",
    doi = "10.1103/PhysRevD.36.3593",
    journal = "Phys. Rev. D",
    volume = "36",
    pages = "3593--3603",
    year = "1987"
}

@article{Campos1993,
    author = "Campos, Antonio and Verdaguer, Enric",
    title = "{Semiclassical equations for weakly inhomogeneous cosmologies}",
    eprint = "gr-qc/9307027",
    archivePrefix = "arXiv",
    reportNumber = "UAB-FT-316",
    doi = "10.1103/PhysRevD.49.1861",
    journal = "Phys. Rev. D",
    volume = "49",
    pages = "1861--1880",
    year = "1994"
}

@book{Calzetta2008,
    author = "Calzetta, Esteban A. and Hu, Bei-Lok B.",
    title = "{Nonequilibrium Quantum Field Theory}",
    doi = "10.1017/9781009290036",
    isbn = "978-1-009-29003-6, 978-1-009-28998-6, 978-1-009-29002-9, 978-0-511-42147-1, 978-0-521-64168-5",
    publisher = "Oxford University Press",
    year = "2009"
}

@article{Frieman1985,
    author = "Frieman, Joshua A.",
    title = "{Particle Creation in Inhomogeneous Space-times}",
    reportNumber = "SLAC-PUB-3855",
    doi = "10.1103/PhysRevD.39.389",
    journal = "Phys. Rev. D",
    volume = "39",
    pages = "389",
    year = "1989"
}

@article{Calmet2021,
    author = "Calmet, Xavier and Casadio, Roberto and Hsu, Stephen D. H. and Kuipers, Folkert",
    title = "{Quantum Hair from Gravity}",
    eprint = "2110.09386",
    archivePrefix = "arXiv",
    primaryClass = "hep-th",
    doi = "10.1103/PhysRevLett.128.111301",
    journal = "Phys. Rev. Lett.",
    volume = "128",
    number = "11",
    pages = "111301",
    year = "2022"
}

@article{Calmet2019,
    author = "Calmet, Xavier and Casadio, Roberto and Kuipers, Folkert",
    title = "{Quantum Gravitational Corrections to a Star Metric and the Black Hole Limit}",
    eprint = "1909.13277",
    archivePrefix = "arXiv",
    primaryClass = "hep-th",
    doi = "10.1103/PhysRevD.100.086010",
    journal = "Phys. Rev. D",
    volume = "100",
    number = "8",
    pages = "086010",
    year = "2019"
}

@book{Vilkovisky:1984,
 author = " Vilkovisky, G. A." ,
    title =   "{ in QUANTUM THEORY OF GRAVITY. ESSAYS IN HONOR OF THE 60TH BIRTHDAY OF BRYCE S. DEWITT, S.M. Christensen, editor}",
    year = "1984"
}

@article{Fradkin:1978,
    author = "Fradkin, E. S. and Vilkovisky, G. A.",
    title = "{Conformal Off Mass Shell Extension and Elimination of Conformal Anomalies in Quantum Gravity}",
    doi = "10.1016/0370-2693(78)90838-9",
    journal = "Phys. Lett. B",
    volume = "73",
    pages = "209--213",
    year = "1978"
}

@article{Frolov:1981,
    author = "Frolov, Valeri P. and Vilkovisky, G. A.",
    title = "{Spherically Symmetric Collapse in Quantum Gravity}",
    doi = "10.1016/0370-2693(81)90542-6",
    journal = "Phys. Lett. B",
    volume = "106",
    pages = "307--313",
    year = "1981"
}

@article{Hamber:2005,
    author = "Hamber, Herbert W. and Williams, Ruth M.",
    title = "{Nonlocal effective gravitational field equations and the running of Newton's G}",
    eprint = "hep-th/0507017",
    archivePrefix = "arXiv",
    reportNumber = "DAMTP-2005-59",
    doi = "10.1103/PhysRevD.72.044026",
    journal = "Phys. Rev. D",
    volume = "72",
    pages = "044026",
    year = "2005"
}

@article{Donoghue:2014,
    author = "Donoghue, John F. and El-Menoufi, Basem Kamal",
    title = "{Nonlocal quantum effects in cosmology: Quantum memory, nonlocal FLRW equations, and singularity avoidance}",
    eprint = "1402.3252",
    archivePrefix = "arXiv",
    primaryClass = "gr-qc",
    reportNumber = "ACFI-T14-004",
    doi = "10.1103/PhysRevD.89.104062",
    journal = "Phys. Rev. D",
    volume = "89",
    number = "10",
    pages = "104062",
    year = "2014"
}

@article{Riegert:1984,
    author = "Riegert, R. J.",
    title = "{A Nonlocal Action for the Trace Anomaly}",
    doi = "10.1016/0370-2693(84)90983-3",
    journal = "Phys. Lett. B",
    volume = "134",
    pages = "56--60",
    year = "1984"
}

@article{Barvinsky:1994,
    author = "Barvinsky, A. O. and Gusev, Yu. V. and Vilkovisky, G. A. and Zhytnikov, V. V.",
    title = "{The Basis of nonlocal curvature invariants in quantum gravity theory. (Third order.)}",
    eprint = "gr-qc/9404061",
    archivePrefix = "arXiv",
    reportNumber = "ALBERTA-THY-14-94",
    doi = "10.1063/1.530427",
    journal = "J. Math. Phys.",
    volume = "35",
    pages = "3525--3542",
    year = "1994"
}

@article{Shapiro:2008sf,
    author = "Shapiro, Ilya L.",
    title = "{Effective Action of Vacuum: Semiclassical Approach}",
    eprint = "0801.0216",
    archivePrefix = "arXiv",
    primaryClass = "gr-qc",
    doi = "10.1088/0264-9381/25/10/103001",
    journal = "Class. Quant. Grav.",
    volume = "25",
    pages = "103001",
    year = "2008"
}

@article{Mottola:2006,
    author = "Mottola, Emil and Vaulin, Ruslan",
    title = "{Macroscopic Effects of the Quantum Trace Anomaly}",
    eprint = "gr-qc/0604051",
    archivePrefix = "arXiv",
    reportNumber = "LA-UR-06-2112",
    doi = "10.1103/PhysRevD.74.064004",
    journal = "Phys. Rev. D",
    volume = "74",
    pages = "064004",
    year = "2006"
}

@book{Fabbri:2005,
    author = "Fabbri, A. and Navarro-Salas, J.",
    title = "{Modeling black hole evaporation}",
    doi = "10.1142/p378",
    isbn = "978-1-86094-527-4, 978-1-86094-722-3, 978-1-78326-038-6",
    publisher = "World Scientific",
    address = "Singapore",
    year = "2005"
}

\end{document}